\documentclass[Journal]{IEEEtran}
\usepackage{amsmath}
\usepackage{amsfonts}
\usepackage{makecell}
\usepackage{amssymb}
\usepackage{graphicx}
\usepackage{epstopdf}
\usepackage[cmintegrals]{newtxmath}
\usepackage{multirow}
\usepackage{cite}
\usepackage{graphicx}
\usepackage{csquotes}

\usepackage{tabularx}
\usepackage{multirow}
\usepackage{rotating}
\usepackage{colortbl}

\newcolumntype{Y}{>{\centering\arraybackslash}X}
\usepackage{makecell}
\usepackage[table,xcdraw]{xcolor} 
\usepackage[acronym,nopostdot]{glossaries}
\usepackage{xurl}
\usepackage[edges]{forest}
\usepackage{balance}
\usepackage{caption}
\usepackage{pifont}
\usepackage{tikz}

\usepackage{dirtytalk}
\usepackage{float} 
\usetikzlibrary{shapes.geometric, positioning, arrows}
\tikzstyle{line}=[draw] 
\tikzstyle{arrow}=[draw, -latex]
 \usepackage[utf8]{inputenc}
     \usepackage{setspace}
\usepackage{amsmath,amssymb,amsfonts}
\usepackage[compact]{titlesec}         
\titlespacing{\section}{0pt}{0pt}{0pt} 
\usepackage{algorithmic}
\usepackage{graphicx}
\usepackage{subcaption}
\usepackage{epstopdf} 
\usepackage{textcomp}
\usepackage{enumitem}

\usepackage{xurl}
\usepackage{xcolor}
\usepackage{colortbl}
\usepackage{etoolbox}
\usepackage{varwidth}

\newcommand{\RN}[1]{%
\textup{\uppercase\expandafter{\romannumeral#1}}
}
\def\BibTeX{{\rm B\kern-.05em{\sc i\kern-.025em b}\kern-.08em
    T\kern-.1667em\lower.7ex\hbox{E}\kern-.125emX}}
\usepackage[acronym,nopostdot]{glossaries}

\makeglossaries
\newacronym{bh}{BH}{Beam hopping}
\newacronym{cdma}{CDMA}{Code division multiple access }
\newacronym{cr}{CR}{Cognitive radio }
\newacronym{csi}{CSI}{Channel state information}
\newacronym{dl}{DL}{Deep Learning}
\newacronym{dra}{DRA}{Dynamic resource allocation}
\newacronym{drl}{DRL}{Deep reinforcement learning}

\newacronym{dvb}{DVB}{Digital video broadcasting}
\newacronym{etsi}{ETSI}{European telecommunications standards institute}
\newacronym{fdma}{FDMA}{Frequency division multiple access}

\newacronym{geo}{GEO}{Geostationary Earth orbit}
\newacronym{gps}{GPS}{Global positioning system}
\newacronym{haps}{HAPS}{High-altitude platform stations }

\newacronym{hts}{HTS}{High throughput satellites}
\newacronym{isl}{ISL}{Inter-satellite links}
\newacronym{leo}{LEO}{Low Earth orbit}
\newacronym{meo}{MEO}{Medium Earth orbit}
\newacronym{mftdma}{MF-TDMA}{Multi-frequency time division multiple access}
\newacronym{ml}{ML}{Machine learning}
\newacronym{modcod}{MODCOD }{Modulation and coding }
\newacronym{nfv}{NFV}{Network function virtualization }
\newacronym{noma}{NOMA}{Non-orthogonal multiple access}
\newacronym{ntn}{NTN}{Non-terrestrial networks}

\newacronym{obp}{OBP}{Onboard processor}
\newacronym{oma}{OMA}{Orthogonal multiple access}
\newacronym{qos}{QoS}{Quality of service}
\newacronym{rf}{RF}{Radio frequency}
\newacronym{rl}{RL}{Reinforcement learning}
\newacronym{rs}{RS}{Rate-splitting }
\newacronym{rsma}{RSMA}{Rate-splitting multiple access}
\newacronym{satcom}{SatCom}{Satellite communication}
\newacronym{satnet}{SatNet}{Satellite network}
\newacronym{sdn}{SDN}{Software-defined networking}
\newacronym{spacenets}{SpaceNets}{Space networks}
\newacronym{sdma}{SDMA}{Spatial division multiple access}

\newacronym{tdma}{TDMA}{Time division multiple access}

\newacronym{3gpp}{3GPP}{3$^{\text{rd}}$ 	\textrm{generation partnership project}}

\newacronym{6g}{6G}{6th generation}

\begin{document}

\markboth{}%
{}
\title{Evolution of High Throughput Satellite Systems: Vision, Requirements, and Key Technologies}

\IEEEoverridecommandlockouts  
\author{\IEEEauthorblockN{Olfa Ben Yahia,~\IEEEmembership{Member,~IEEE,} Zineb Garroussi,~\IEEEmembership{Member,~IEEE,} Olivier Bélanger,~\IEEEmembership{Student Member,~IEEE,} Brunilde~Sansò,~\IEEEmembership{Senior Member,~IEEE,} Jean-François Frigon,~\IEEEmembership{Senior Member,~IEEE,} Stéphane Martel, Antoine~Lesage-Landry,~\IEEEmembership{Member,~IEEE,} and Gunes~Karabulut~Kurt,~\IEEEmembership{Senior Member,~IEEE}}

\thanks{O. Ben Yahia, Z. Garroussi, O. Bélanger, B. Sansò, J. Frigon, A. Lesage-Landry, and G. Karabulut-Kurt are with the Department of Electrical Engineering, Polytechnique Montréal, Montréal, QC H3T 1J4, Canada.
Emails:\{olfa.ben-yahia, zineb.garroussi, olivier.belanger, brunilde.sanso, j-f.frigon, antoine.lesage-landry, gunes.kurt\}@polymtl.ca}
\thanks{S. Martel is with Satellite Systems, MDA, 21025 Trans-Canada Hwy Sainte-Anne-de-Bellevue, Qc H9X 3R2. Email: Stephane.Martel@mda.space} 
\thanks{
This work is supported by MDA, CRIAQ, MITACS, and NSERC.}}

\maketitle

\begin{abstract}

High throughput satellites (HTS), with their digital payload technology, are expected to play a key role as enablers of the upcoming 6G networks. HTS are mainly designed to provide higher data rates and capacities. Fueled by technological advancements including beamforming, advanced modulation techniques, reconfigurable phased array technologies, and electronically steerable antennas, HTS have emerged as a fundamental component for future network generation. This paper offers a comprehensive state-of-the-art of HTS systems, with a focus on standardization, patents, channel multiple access techniques, routing, load balancing, and the role of software-defined networking (SDN). In addition, we provide a vision for next-satellite systems that we named as extremely-HTS (EHTS) toward autonomous satellites supported by the main requirements and key technologies expected for these systems. The EHTS system will be designed such that it maximizes spectrum reuse and data rates, and flexibly steers the capacity to satisfy user demand. We introduce a novel architecture for future regenerative payloads while summarizing the challenges imposed by this architecture.

\end{abstract}


\glssetwidest{HAPS-SMBS}
\hspace{1cm}\printglossary[style=alttree,type=\acronymtype,title=\textbf{Abbreviations},nogroupskip, nonumberlist]

\IEEEpeerreviewmaketitle

\section{Introduction}
\label{sec:Introduction}


%

The growing demand for novel satellite services and systems is fueling the development of innovative approaches that depart from traditional single-beam designs. Instead, there is a shift toward advanced multi-beam implementations, aiming to enhance network capacity. High throughput satellites (HTS) with multiple beams are a new generation of satellites designed to provide higher bandwidth and capacity than traditional satellites~\cite{zhang2023high}. 

\begin{figure*}
  \centering
\includegraphics[width=1\textwidth, height=8in,keepaspectratio]{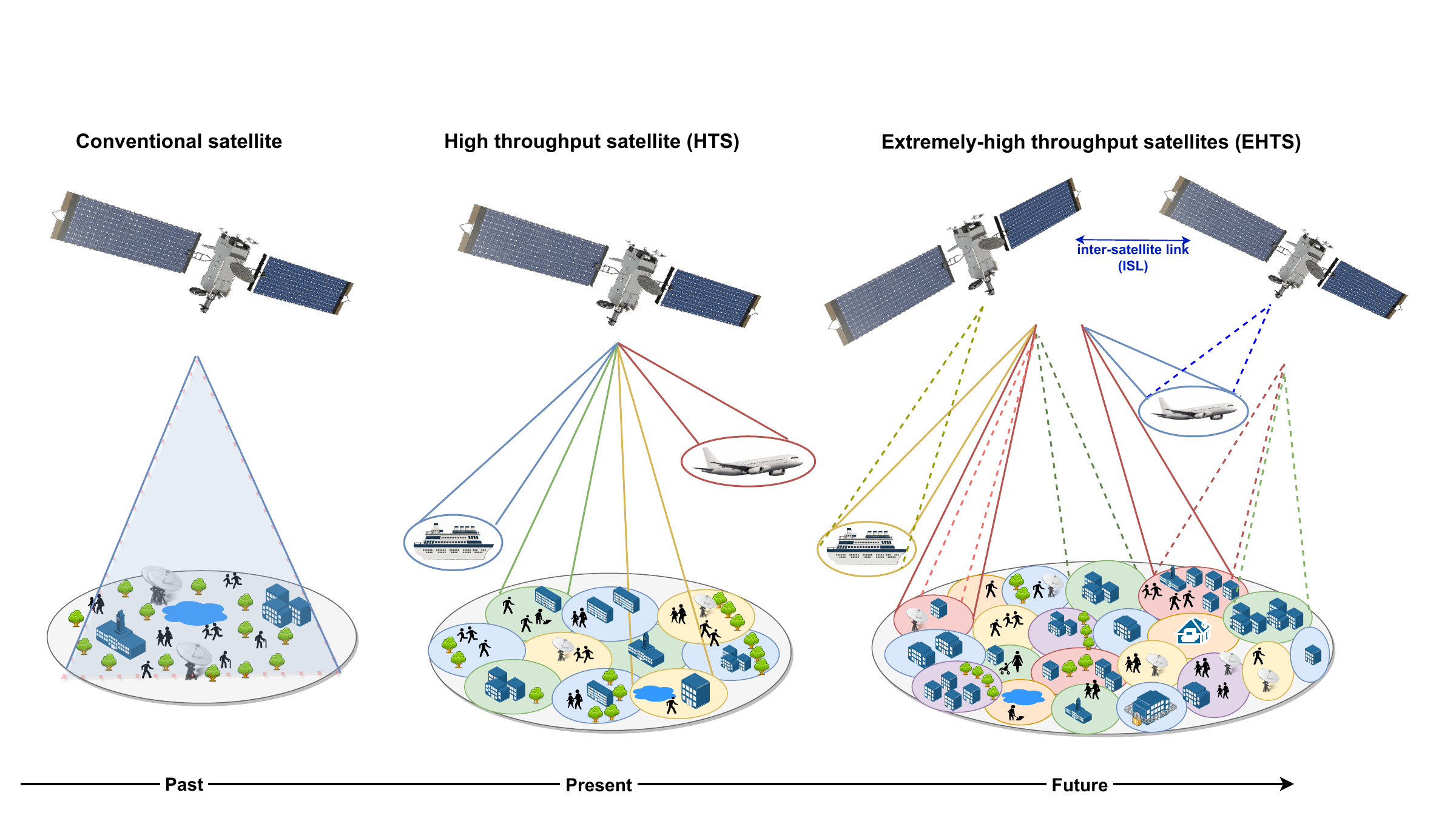}
   \caption{Evolution of SatCom from single-beam satellite to EHTS.}
  \label{figure_ehts}
\end{figure*}

These satellites are now integrating cutting-edge digital payload technologies to cater to a variety of markets and evolving applications. Their payloads possess the capacity to reconfigure and perform several functions such as modifying beam coverage, managing satellite resources, and adjusting the radio frequency (RF) power distribution dynamically based on traffic demands~\cite{yun2013future, braun2021processing}. Nonetheless, enhanced management strategies need to be utilized to effectively leverage the potential of these digital payloads. According to~\cite{abdu2023demand}, the payload operates signals using multiple digital processors. This is because a single processor can manage a limited number of beams to minimize signal processing complexity and delay from sequential task processing. Also, signals with the same carrier frequency need distinct processors to prevent interference. Therefore, a proper allocation of signal carrier bandwidth between multiple processors is necessary. 

With an ongoing data-driven world, emerging use cases, and the increasing demand for higher throughputs and capacities, we envision the future with more evolved and autonomous satellites, that we term extremely-HTS (EHTS). These satellites will provide more flexible and adaptive resource management with extremely-high throughput satellites as depicted in Figure \ref{figure_ehts}. Later in the paper, we provide a detailed exploration of this vision.


\subsection{Research Gap and Current Challenges in Modern Satellites Systems}
For future satellites, including HTS and beyond, it is expected that the payload architecture can be modelled as multiple modem banks or multiple packet forwarders and a packet routing and switching engine ~\cite{hindin2019technical}. With full packet regeneration, repackaging, and prioritization, all traffic within the satellites is fully routable between various user and gateway modem banks. A significant challenge imposed by the type of modem bank network is managing the high number of crossovers between successive modem banks, making suitable routing decisions for each demand and efficiently allocating resources within the payload system. This challenge directly affects the performance and efficiency of multi-beam satellite systems. There is a need for faster and more effective resource allocation strategies as existing research has limitations when optimizing the data flow routing within the HTS system. In the recent years, researchers have devoted considerable effort to investigating various methods of resource allocation in HTS communication. This includes optimizing factors such as time slots, spectrum, and power, among others.
However, effectively routing flows of data within the satellite's multiple modem banks to reach the appropriate egress beam is a critical challenge that has not been adequately addressed in the existing literature.

\begin{figure}[!h]
  \centering
   \includegraphics[width=1\textwidth, height=7in,keepaspectratio]{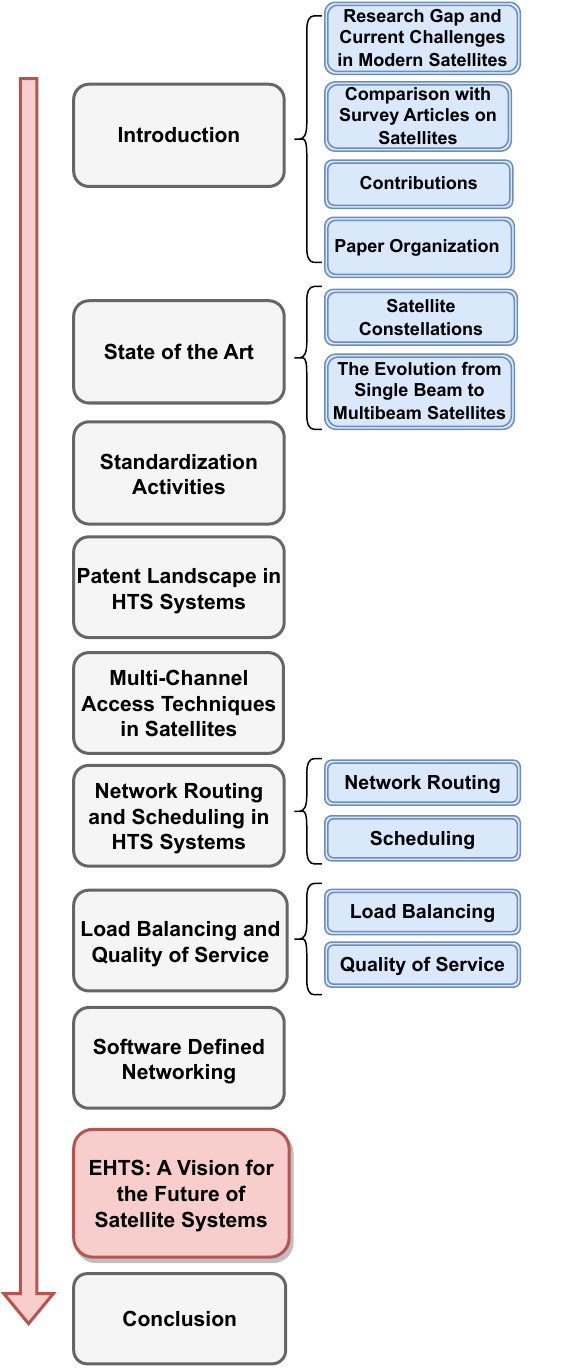}
  \caption{Organization of the paper.}
  \label{figure1}
\end{figure}


\begin{table*}
\small
\renewcommand{\arraystretch}{1.3}
\centering
\caption{Overview of Survey Articles on Satellite Systems. (N: Not covered, L: Low coverage, M: Medium coverage, H: High coverage)}
\label{tab:comparison}
\begin{tabular}{|c|c|c|c|c|c|c|c|c|c|} 
    \hline
    & \multicolumn{1}{c|}{\textbf{Section \ref{sec:Standardization}}} & \multicolumn{1}{c|}{\textbf{ Section \ref{sec:Patents} }} & \multicolumn{1}{c|}{\textbf{Section  \ref{multi-ch-access}}} & \multicolumn{2}{c|}{\textbf{Section \ref{sec:Routingandsch}}} & \multicolumn{2}{c|}{\textbf{Section  \ref{sec:LoadBalancing_QOS}}} & \multicolumn{1}{c|}{\textbf{Section \ref{sec:SDN}}} & \multicolumn{1}{c|}{\textbf{Section \ref{sec:EHTS}}} \\
    \hline
 \hline
\parbox[t]{2mm}{\multirow{5}{*}{\rotatebox[origin=c]{90}{\textbf{Reference}}}} &  
\parbox[t]{2mm}{\rotatebox[origin=c]{90}{Standardization}} &
\parbox[t]{2mm}{\rotatebox[origin=c]{90}{Patents}} &
\parbox[t]{2mm}{\rotatebox[origin=c]{90}{Multichannel Access}} &
\parbox[t]{2mm}{\rotatebox[origin=c]{90}{Network Routing}} & {\rotatebox[origin=c]{90}{Scheduling}} &
\parbox[t]{2mm}{\rotatebox[origin=c]{90}{Load Balancing }} & {\rotatebox[origin=c]{90}{Quality of Service}} &
\parbox[t]{2mm}{\rotatebox[origin=c]{90}{Software Defined Networking}} &
\rotatebox[origin=c]{90}{\parbox{5cm}{ \centering EHTS: A Vision for the Future \\ of Satellite Systems}} \\
  
    \hline
    \textbf{Kodheli \textit{et al.} ~\cite{9210567}} & \cellcolor{yellow}M & \cellcolor{red}N & \cellcolor{orange}L & \cellcolor{orange}L & \cellcolor{red}N  & \cellcolor{red}N  & \cellcolor{red}N & \cellcolor{green}H & \cellcolor{red}N \\
    \hline
    \textbf{Abdelsadek \textit{et al.}~\cite{9982444}} & \cellcolor{green}H & \cellcolor{red}N & \cellcolor{red}N & \cellcolor{red}N  & \cellcolor{red}N & \cellcolor{red}N & \cellcolor{red}N & \cellcolor{green}H & \cellcolor{red}N \\  
    \hline
    \textbf{Rinaldi \textit{et al.}~\cite{9193893}} & \cellcolor{yellow}M  & \cellcolor{red}N  & \cellcolor{red}N  & \cellcolor{red}N  & \cellcolor{red}N  & \cellcolor{red}N  & \cellcolor{orange}L  & \cellcolor{orange}L & \cellcolor{red}N  \\
    \hline
    \textbf{Guidotti \textit{et al.}~\cite{8626457}} & \cellcolor{yellow}M  & \cellcolor{red}N & \cellcolor{red}N  & \cellcolor{red}N & \cellcolor{orange}L  & \cellcolor{red}N & \cellcolor{red}N & \cellcolor{red}N & \cellcolor{red}N  \\
    \hline
    \textbf{Radhakrishnan \textit{et al.}~\cite{7466793}} & \cellcolor{red}N & \cellcolor{red}N & \cellcolor{red}N & \cellcolor{green}H & \cellcolor{red}N & \cellcolor{red}N & \cellcolor{red}N & \cellcolor{red}N & \cellcolor{red}N \\
     \hline
    \textbf{Liu \textit{et al.}~\cite{liu2018space}} & \cellcolor{orange}L & \cellcolor{red}N & \cellcolor{red}N & \cellcolor{green}H & \cellcolor{red}N  & \cellcolor{orange}L & \cellcolor{yellow}M & \cellcolor{green}H & \cellcolor{red}N\\
     \hline
    \textbf{Burleigh \textit{et al.}~\cite{burleigh2019connectivity}} & \cellcolor{red}N & \cellcolor{red}N & \cellcolor{red}N & \cellcolor{orange}L & \cellcolor{red}N & \cellcolor{red}N & \cellcolor{red}N & \cellcolor{yellow}M & \cellcolor{red}N \\
    \hline
     \textbf{Azari \textit{et al.}~\cite{9861699}} & \cellcolor{yellow}M & \cellcolor{red}N & \cellcolor{red}N & \cellcolor{orange}L & \cellcolor{orange}L & \cellcolor{red}N & \cellcolor{red}N & \cellcolor{yellow}M & \cellcolor{red}N \\
    \hline
    \textbf{This paper} & \cellcolor{green}H & \cellcolor{green}H & \cellcolor{green}H & \cellcolor{green}H & \cellcolor{green}H & \cellcolor{green}H & \cellcolor{green}H  & \cellcolor{green}H & \cellcolor{green}H \\
    \hline
\end{tabular}
\end{table*}

\normalsize
\subsection{Comparison with Survey Articles on Satellites}

When comparing our work to surveys of the state of the art in satellite communication (SatCom) or satellite network (\acrshort{satnet})~\cite{9210567, 9982444, 9193893, 8626457, 7466793, 9861699,liu2018space,burleigh2019connectivity}, it is clear that the majority of the literature highlights promising research areas including network architecture, standardization, network design, various use cases, and open research directions. For instance, the authors of~\cite{9982444} present a broad vision for future space networks (\acrshort{spacenets}) that takes into account developments in a number of related sectors. Other works consider analyzing the \acrfull{3gpp} non-terrestrial networks (\acrshort{ntn}) characteristics and their prospective uses. Our work differs from previous surveys, nevertheless, in that we pay close attention to HTS. There is currently no literature that clearly describes the state of the art for HTS or gives a thorough overview of the networking aspects of HTS, an area we seek to fill. It is noteworthy to mention that reference~\cite{liu2018space} delves into the networking aspect including resource management and allocation for space-air-ground integrated network.

\subsection{Contributions}

It is expected that traditional routing algorithm-based load balancing, optimization problem, scheduling, and quality of service (QoS) policies alongside monitoring and software-defined network (SDN) can be further investigated, improved, and adapted to future regenerative payloads. Thereafter, this review study is methodically organized to provide a comprehensive understanding of the fundamental components of HTS systems while also providing an important contribution by filling the substantial gap in the existing literature. Our research contributes to the current understanding of HTS by bringing together multiple perspectives into a comprehensive assessment of topics such as standards, patents, multiple channel access techniques, routing, load balancing, and the important role of SDN. In addition, in this paper, we discuss the networking aspects of a regenerative satellite constellation. Furthermore, we distinguish between HTS systems and next-satellite systems as EHTS. Lastly, we also propose a new architecture for future payloads based on multiple modem banks and present the challenges imposed by this architecture.

\subsection{Paper Organization}
The purpose of this survey is to present the potential solutions and to address the current challenges of modern satellites, while also highlighting the unresolved problems existing in the current literature. Along this direction, Section \ref{Sect:Motiv} presents the state of the art and outlines the main motivations behind this work. Section \ref{sec:Standardization} delves into the development and standardization of satellite systems which may have an indirect impact on HTS evolution. Section \ref{sec:Patents} presents an overview of the HTS patent landscape. In Section \ref{multi-ch-access}, we summarize the existing multiple-channel access techniques in SatCom. In Subsection \ref{sec:Routing}, we review existing routing algorithms for single-beam satellites considering different satellite architectures, recognizing a noticeable gap in the current literature regarding routing mechanisms within HTS systems. Additionally, we shed light on the various scheduling mechanisms, examining their impacts on HTS performance in Subsection \ref{sec:Scheduling}. Section \ref{sec:LoadBalancing_QOS} then turns to load balancing and QoS in HTS, discussing their roles in enhancing HTS efficiency. Section \ref{sec:SDN} explores the significant role of SDN in  HTS, and examines the current SDN technologies in use and their potential impact on future HTS. 
In Section \ref{sec:EHTS}, we present our vision for future EHTS systems and their main key features.
Finally, Section \ref{sec:Conclusion} concludes the paper, summarizing our findings and highlighting their significance. The overall structure of the paper is summarized in Figure \ref{figure1}.

\section{State of the Art}
\label{Sect:Motiv}
In recent years, there has been a significant increase in demand for data connectivity, driven by the growth of digital technologies and the increasing number of connected devices. 
Given the rising demand for new services and in light of upcoming technological advancements, research into post-5G solutions and 6G technologies for the 2030 era is already gaining significant momentum in both academia and industry~\cite{zhang20196g,latva2020key}. 6G is expected to introduce groundbreaking communication technologies, such as those operating at Terahertz and optical frequencies, along with innovative network architectures~\cite{giordani2020satellite}. Research into 6G is also concentrating on advancing non-terrestrial networks to deliver three-dimensional coverage, supplementing ground-based infrastructures with aerial platforms including satellites and high altitude platform stations (HAPS) systems~\cite{9380673}. As defined by the \acrshort{3gpp}~\cite{3GPP2020}, \acrshort{ntn} is a network that operates either partially or entirely through a spaceborne platform or airborne vehicle, for the purpose of communication. This could include vehicles in geostationary or non-geostationary Earth orbit or airborne vehicles such as \acrshort{haps} and unmanned aerial vehicles (UAVs). The defining characteristic of NTN is their ability to deliver connectivity to regions that are otherwise inaccessible or remote locations where constructing a terrestrial infrastructure would entail significant investment~\cite{azari2022evolution}. Nonetheless, within the 3GPP NTN framework, satellites have been the primary focus, while HAPS are seen as a specific application within a satellite system. While UAVs form a portion of airborne networks, their standardization follows a distinct path within 3GPP.

For decades, SatCom has been largely limited to voice and TV broadcasting services, meteorological Earth observation, as well as universal communication and navigation systems~\cite{sweeting2018modern}. However, recent technological progress in space communication has led to the evolution of SatCom systems toward SpaceNets, offering a more efficient and cost-effective approach to delivering high-speed data connectivity to remote and underserved areas through the deployment of satellite constellations. The Future SpaceNet is envisioned as a network that will seamlessly connect various components such as space nodes (SNs), Earth orbit satellite networks (EOSN), and space network terminals (SNTs). Moreover, it will serve as the bridge connecting these components to other networks like the Internet, aerial networks, and cellular networks~\cite{9982444}. Specifically, SpaceNets with networking capacities will revolutionize the way we communicate, navigate, and gather information about our planet. A variety of satellite systems exist, each with their own unique characteristics and applications including low Earth orbit (LEO), medium Earth orbit (MEO), and geostationary Earth orbit (GEO). In the following discussion, we will delve into each of these satellite types in more detail.

\subsection{Satellite Constellations}

\acrshort{geo} satellites are placed in orbits that are approximately 36,000 kilometers above the Earth's surface. These satellites are primarily used for SatCom, such as television broadcasting and Internet services. Because GEO satellites are in a fixed position relative to the Earth, they can provide continuous coverage to a specific geographic region. However, because GEO satellites are further away from Earth, they have a longer communication latency~\cite{centenaro2021survey}.

\acrshort{meo} satellites are placed in orbits that are typically between 2000 and 36,000 kilometers above the Earth's surface. These satellites are primarily used for navigation systems, such as the Global Positioning System (GPS). \acrshort{meo} satellites have a longer orbital period than LEO satellites, which makes them easier to maintain a continuous connection. However, because \acrshort{meo} satellites are further away from  Earth, they have a longer communication latency~\cite{centenaro2021survey}.

\acrshort{leo} satellites orbit the Earth at an altitude ranging from 160 to 2000 kilometers and are utilized for various purposes such as Earth observation, scientific research, and SatCom. Due to their closer proximity to Earth, LEO satellites offer a lower communication latency that is advantageous for applications that require real-time communication. Nevertheless, because \acrshort{leo} satellites have a shorter orbital period, they have to be in a perpetual state of motion, which can pose difficulties in maintaining a consistent connection~\cite{9210567}. Recently, there has been a shift towards SpaceNets with large satellite constellations~\cite{abdelsadek2022}. This has given rise to mega constellations, with SpaceX planning to possibly deploy 12,000 satellites under its Starlink project. Similar initiatives are being considered by OneWeb and Blue Origin, all with the ambition to offer broadband connectivity on a global scale~\cite{osoro2021techno}. LEO systems allow for the utilization of high-frequency bands like Ku, Ka, Q, and V, which offer larger bandwidths compared to those for GEO satellites~\cite{dvovrak2020atmospheric, kodheli2019link}, thereby enabling higher capacities for users. Furthermore, relative to GEO, the intricacy and cost per satellite in an LEO system is lower, and system redundancy can be consistently enhanced without causing disruption to the overall system. From an economic perspective, LEO systems offer greater scalability than other systems in terms of capacity augmentation. This is because a constellation can be effortlessly expanded without causing any interruption to the existing broadband services~\cite{gaber20205g}. This has made them desirable for a variety of applications, including high-speed trains~\cite{wang2020content}, aircraft~\cite{sturdivant2018systems}, maritime~\cite{InmarsatMaritime, StarlinkMaritime}, and navigation~\cite{iannucci2020economical,ma2020hybrid}. Overall, the most significant benefit of LEO systems and satellites is their capability to cater to remote areas~\cite{cello2015hotsel,agnelli2014satellite}. This is particularly applicable in serving extreme topographies such as cliffs, valleys, steep slopes, and geologically disaster-prone areas, where the implementation of terrestrial networks can be prohibitively expensive due to engineering complexities and cost considerations.

\subsubsection{Payload architectures: transparent vs. regenerative }

According to the 3GPP~\cite{3GPP2020}, a satellite can be equipped with one of the two following payload configurations:

\begin{itemize}
    \item The bent-pipe payload is responsible for handling RF filtering of the uplink signal, frequency conversion, and amplification before transmitting it back to the downlink.
    \item The regenerative payload, on the other hand, in addition to performing the same tasks as the bent-pipe payload, possesses abilities like demodulation and decoding, switching and/or routing, and coding and/or modulation.

\end{itemize}

As an alternative to bent-pipe payloads, regenerative space platforms embrace the idea of on-board processing. On the one hand, they provide a significant extra benefit, notably the capacity to carry out routing directly on the satellite. Therefore, inter-satellite links (ISL) can be used to direct data after it has been demodulated on the satellite and directed to a variety of locations~\cite{jia2023network}. This enables data flow inside the constellation to be routed across high-rate optical or RF links between satellites on the same plane or across separate planes. For instance, based on elements like cost-efficiency or legal restrictions, satellites that service aviation and maritime traffic can route the data through ISLs to the appropriate terrestrial gateway. In addition, regenerative payloads have the advantage of enabling single-hop mesh connections between conventional terminals. In contrast, mesh communication in bent-pipe systems often requires dual-hop communication, via the ground gateway~\cite{shachamboard}. On the other hand, regenerative systems are frequently more complicated than bent-pipe transponders, despite the fact that they have several advantages, such as a completely regenerative payload with functions for all protocols (such as routing and queuing). Implementing this solution would necessitate more complex control and the development of novel schemes.

   \begin{table*}[!t]
\centering
\caption{Overview of Key Features and Advantages of HTS }
\begin{tabular}{|l|p{12cm}|}
\hline
\textbf{Requirement} & \textbf{Description} \\ \hline
High data rate and wide coverage & \acrshort{hts}  are designed to provide significantly more data throughput than traditional satellites, typically in the range of several hundred Gbps, and cover large geographic areas, often spanning entire continents or even the entire globe. This allows for faster and more reliable data transmission, even in remote and underserved areas where traditional infrastructure may not be available~\cite{perez2019signal}. \\ \hline

Beam hopping (BH) &  BH is considered a key technological enabler for satellite systems by offering significant flexibility to handle unpredictable and time-variant traffic requests within a satellite's coverage area~\cite{anzalchi2010beam}. This technology allows all available satellite resources to be used to serve a specific subset of beams for a certain duration, dwelling just long enough to meet the demand in each beam~\cite{lei2020beam, chen2022next}.\\ \hline
Quality of service (QoS) & \acrshort{hts} systems can cater to user demands for high-quality, high data-rate services, essential for emerging applications like 3D, 4K, and ultra high definition TV~\cite{muhammad2016qos}. \\ \hline
Frequency reuse & Multi-beam coverage allow for the repeated use of the same frequency/polarization sub-band, significantly enhancing the utilization of bandwidth by an equivalent amount~\cite{6400146}.\\ \hline
Spot beams & Contrary to conventional satellite technologies that use a single (global) beam, HTS systems can utilize a substantial number of spot beams to cover the same area. It is possible to employ tens or even hundreds of beams, typically with a reuse factor of four~\cite{6184256}. \\ \hline
Digital processing & The digital transparent processor or mesh processor on a multi-beam communication satellite offer numerous benefits. It enhances multimedia experiences by reducing transit delays, provides secure communications, and offers disaster-resilient connectivity. It also brings operational flexibility, conserves bandwidth, eliminates hub-related costs, and excels in high-traffic mobile satellite services due to superior antenna performance capabilities~\cite{shah2015onboard}. \\ \hline
Multiple frequency bands & HTS often operate in multiple frequency bands, including both Ka-band and Ku-band, to provide maximum flexibility and capacity~\cite{perez2019signal}. \\ \hline
Advanced antennas & \acrshort{hts} often use advanced antennas, including new reconfigurable phased array technologies and electronically-steerable antennas, to improve performance and increase capacity~\cite{chaloun2022electronically}. \\ \hline
Higher transmit/receive gain & A narrower beam, due to its higher directivity and gain, result in augmented power for both transmission and reception. This facilitates the use of smaller user terminals and enables the application of higher-order modulations. Consequently, it allows for increased data transmission rates per unit of the orbital spectrum~\cite{swinford2015high}. \\ \hline
Security & Spot beam design and advanced digital payloads significantly enhance the resistance to interference and jamming. Features such as onboard power monitoring, notch filtering, and frequency hopping modems provide low probability-of-intercept and interference-mitigation capabilities. These elements, combined with the ability to reroute signals, improve user protection in contested environments~\cite{intelsatgeneral2020}.\\ \hline
\end{tabular}
\label{table:hts-requirements}
\end{table*}

\subsection{The Evolution from Single-Beam to Multi-Beam Satellites }

 \begin{table*}[!t]
    \caption{HTS Service Providers (Alphabetical order)}
    \small
    \centering
        \begin{tabular}{|p{1.5cm}|p{1.2cm}|p{5cm}|p{2.5cm}|p{2.5cm}|p{1.5cm}|p{1.6cm}|}
        \hline
        \textbf{Company} & \textbf{Country}& \textbf{Notable Features} & \textbf{Project example} & \textbf{Data Rate (Gbps,~Up To)} & \textbf{Frequency Band} & \textbf{Orbital Altitude}\\ \hline
        

        Amazon \cite{foust_amazon_2020, Kuiper2019} & USA & High data rate, multi-beam technology, Advanced DSP, Global coverage & Kuiper & TBA & Ka & LEO \\ \hline
        
        

        China Satcom~\cite{jones2023china} & China & Satellite communication services, Operations of various satellite constellations & Zhongxing-26 (ChinaSat-26) & 100+ & Ka & GEO \\ \hline
        
        Eutelsat~\cite{eutelsat_airaccess}& France & High data rate, Multi-beam technology, Advanced DSP, Resilience and security features, On-board processing & KA-SAT & 90 & Ka & GEO  \\ \hline
        
        Hughes Network Systems~\cite{hughes_jupiter} & USA & High data rate, Multi-beam technology, Advanced DSP, Resilience and security features, Electric propulsion &  JUPITER~3 (EchoStar XXIV)  & 500+ & Q, V, Ka & GEO  \\ \hline
        
        Intelsat~\cite{intelsat_epicng} & USA & Multi-beam technology, Advanced DSP, Resilience and security features, Global coverage & Intelsat Epic & 25-60 & Ku, C, Ka & GEO \\ \hline

        ISRO \cite{isro2018,wion2022}  & India & Multi-beam, Multi-band technology, based on enhanced I3K bus, High frequency technology & \begin{tabular}[t]{@{}l@{}}GSAT-11 \\ GSAT-29\end{tabular} & \begin{tabular}[t]{@{}l@{}}1 \\ (Potential: 10-100)\end{tabular}  & Ka, Ku & GEO \\ \hline
        
        OneWeb~\cite{griffin_oneweb_2022} & UK & High data rate, multi-beam technology, Laser communication technology, Advanced DSP, Resilience and security features, On-board processing, On-orbit servicing and repair capabilities & OneWeb Constellation & 7.2 & Ka, Ku & LEO \\ \hline
        
        SES~\cite{ses_2021, spaceflight101_o3b_2023} & Luxemb. & High data rate, Multi-beam technology, Advanced DSP, Resilience and security features, On-board processing & \begin{tabular}[t]{@{}l@{}}O3b mPOWER \\ SES-17\end{tabular}  & \begin{tabular}[t]{@{}l@{}}10 \\ 400\end{tabular} & Ka & \begin{tabular}[t]{@{}l@{}}MEO \\ GEO \end{tabular} \\ \hline
        
        
        SpaceX~\cite{pekhterev_starlink_bandwidth} & USA & Mass-manufacturing capability, Low-cost launch integration, On-orbit satellite servicing potential, Phased array broadband antennas, Redundant satellite design & Starlink  & $\sim 20$ & Ka, Ku, E & LEO \\ \hline
        
        Telesat~\cite{telesat_2022, rainbow2023telesat} & Canada & High data rate, Multi-beam technology, Laser communication technology, Advanced DSP, Resilience and security features, On-board processing, On-orbit servicing, and repair capabilities & \begin{tabular}[t]{@{}l@{}}Lightspeed \\ Telstar 19 Vantage\end{tabular} & \begin{tabular}[t]{@{}l@{}}50 \\ 31 \end{tabular} & \begin{tabular}[t]{@{}l@{}}Ka \\ Ka, Ku\end{tabular} & \begin{tabular}[t]{@{}l@{}}LEO \\ GEO\end{tabular}  \\ \hline
        
        
        Viasat~\cite{baumgartner_viasat_2023} & USA & High data rate, Multi-beam technology, Advanced DSP, Resilience and security features, On-board processing, Electric propulsion & Viasat-3 & 1000 & Ka & GEO   \\ 
        \hline
        \end{tabular}
        \label{Service Providers}
\end{table*}

The growing demand for novel satellite services and systems is fueling the development of innovative approaches that depart from traditional single-beam designs. Instead, there is a shift towards advanced multi-beam implementations, aiming to enhance network capacity. HTS with multiple beams are a new generation of satellites designed to provide higher bandwidth and capacity than traditional satellites. Through the integration of sharply concentrated beams and frequency reuse, HTS are being engineered and constructed to deliver a throughput ranging from hundreds of gigabits per second (Gbit/s) to over one terabit per second (Tbit/s) from each orbital position~\cite{itu-r2019key-elements}.
These satellites use advanced digital signal processing, multiple concentrated spot beams, wideband payloads, increased frequency re-use, and higher frequency bands to provide significant high-speed data connectivity over wide geographical areas~\cite{itu-r2019key-elements}. These systems are versatile, catering to a range of applications such as the distribution of content via broadcast and multicast. Additionally, they contribute to narrowing the digital gap by providing high-speed, high-capacity services accessible anytime and anywhere. The main key technologies of HTS are enhanced interference mitigation, dynamic beam formation, adaptable beam hopping (\acrshort{bh}), onboard processing, and an even more extensive reuse of frequency.

Currently, numerous operators are engaged in upgrading their constellations to offer improved capabilities such as higher \acrshort{rf} power, enhanced functionality, and increased frequency reuse. These upgrades are being accomplished through the adoption of \acrshort{hts} or very high throughput satellites (VHTS) technologies~\cite{perez2019signal}. In this context, Amazon's Kuiper System will offer high-speed, low-latency satellite broadband services through a fleet of 3,236 Ka-band satellites located at orbital altitudes of 590 km, 610 km, and 630 km. The system leverages cutting-edge communication antennas, sub-systems, and semiconductor technology to deliver cost-effective broadband services for both consumer and enterprise usage. This includes IP transit, carrier-grade Ethernet, and wireless backhaul traffic. Designed to optimize spectrum reuse and efficiency, the Kuiper system has the flexibility to steer capacity according to regional customer demand. Furthermore, it benefits from Amazon's terrestrial networking infrastructure, enabling it to provide secure, high-speed, low-latency broadband services to its customers~\cite{hindin2019technical}. Table \ref{Service Providers} offers a comparative perspective on different HTS service providers.

Additionally, there are several promising techniques and paradigms being developed to enhance the flexibility and dynamism of future \acrshort{hts} mainly reconfigurable payloads, SDN, and network function virtualization (\acrshort{nfv}). These advancements aim at increasing the flexibility and dynamic capabilities of HTS systems, allowing for more efficient and adaptable SatNets~\cite{sharma2020system}. In addition, the onboard processor (\acrshort{obp}) in satellites extends the network and service capabilities, going beyond the traditional bent pipe satellite. The main key features of an \acrshort{obp} include~\cite{campanella1987board}:
\begin{enumerate}
    \item On-board demultiplexing and demodulation of individual uplink carriers using digital frequency division multiple access (FDMA), time division multiple access (TDMA), or a combination like multiple frequency TDMA (MF/TDMA).
    \item Rerouting of individual channels from a multitude of earth stations through multiple regional and spot beams into different downlink transmission channels.
    \item Remodulation of rearranged channels onto downlink carriers within multiple regional and spot beams, using TDMA or MF/TDMA for hopping beams, and FDMA or TDMA for fixed beams.
\end{enumerate}
Therefore, \acrshort{obp} allows for data processing and adjustment directly on the satellite, diminishing the necessity for regular interaction with terrestrial stations.

\vspace{0.4cm}

To gain further insight into HTS, some key characteristics are outlined in Table \ref{table:hts-requirements}. From Table \ref{table:hts-requirements}, we can observe that in terms of requirements, HTS are designed to deliver high data rates over wide coverage areas. This is facilitated through several key features like multi-beam technology, advanced antennas, digital processing, and the use of multiple frequency bands. Furthermore, a focus on QoS, security, and frequency reuse indicates that these systems are designed with robustness and efficiency. We also notice the emphasis on high transmit/receive gain to increase the power, to boost data transmission rates. In summary, HTS technology, characterized by innovative technological solutions, drives toward higher performance and efficiency.


\section{Standardization Activities}
\label{sec:Standardization}

\acrshort{hts} have become an essential component of the global communications infrastructure. In order to ensure interoperability and efficient utilization of these satellite systems, standardization is necessary. At this point, the active participation of several standardization organizations, including the \acrfull{etsi} and the \acrshort{3gpp}, becomes critical. Their dedication to establishing and implementing standards is vital for the continued efficient operation of the global SatNet. These organizations play an instrumental role in promoting compatibility across various systems and ensuring the effective use of this invaluable communications asset.

A crucial contributor to the standardization of HTS communication is the digital video broadcasting (\acrshort{dvb}) project, published by the ETSI. DVB has been instrumental in the development of several standards for satellite broadcasting and communication. This includes DVB-S, explicitly designed for satellites, its follow-up DVB-S2, and the more recently introduced DVB-S2X~\cite{ETSI_DVB-S2}. The pivotal characteristic of DVB-S2 is its capacity to dynamically adapt modulation and coding based on the channel quality each user encounters. The standard lays out the rules for the physical layer and data link layer of the communication link, which are layers 1 and 2 in the OSI model. It instructs on how to convert one or more baseband digital signals into an RF signal that is compatible with the satellite channel's characteristics~\cite{etsi2014302}. In 2015, ETSI released amendments to the DVB-S2 standard, identified as DVB-S2X~\cite{etsi2015302}. DVB-S2X can be seen as an enhancement of the previous DVB-S2, representing an evolution of the second-generation digital satellite broadcasting system. This standard introduces several improvements and new features, which include higher modulation order, finer gradation of modulation and coding (\acrshort{modcod}) schemes, superframes specifically to support BH, channel bonding, additional scrambling options, and a variety of new operational modes to suit different applications and environments~\cite{ETSI_DVB-S2}. In particular, the DVB-S2X's increased efficiency makes it particularly relevant for HTS systems. Furthermore, ETSI formulated a set of standards referred to as DVB-return channel via satellite (DVB-RCS)~\cite{etsi101}. These standards specify the air interface requirements for bidirectional communication involving broadband very small aperture terminal (VSAT) terminals~\cite{itu-r2010multi-carrier}. Specifically, the DVB-RCS2 standard outlines the principles for coding, modulation, and multi-user access schemes utilized for a return link that numerous users share via a satellite channel. It expands the adaptive coding and modulation (ACM) functionality to the return link and provides guidelines for the transfer of forward-link quality data to the gateway~\cite{braun2012satellite}. The DVB standards provide guidelines on broadcasting services including TV and radio, data services, and professional applications like contribution links and satellite newsgathering. These DVB standards play a crucial role in facilitating interoperability, efficiency, and service quality in satellites. This helps create a well-functioning global SatNet ecosystem that effectively meets various user demands.

Another key enabler in the telecommunication industry, 3GPP is playing a significant role in setting standards, and has started integrating \acrshort{ntn} into their framework~\cite{9579443}, demonstrating active involvement in SatNet standardization.

Satellites were first referenced as an extension to terrestrial networks in a 5G deployment scenario in the 3GPP TR 38.913 Release 14. The objective was to provide 5G communication services in areas lacking terrestrial coverage, as well as to support services that could be more efficiently accessed via satellite systems. These include broadcasting and delay-tolerant services~\cite{darwish2022leo}. The growing interest in integrating satellites with 5G led 3GPP to define NTNs. Although NTNs incorporate other aerial systems, like HAPS, the 3GPP community views satellites as the primary case, and other aerial systems as special instances of satellites. Starting with 3GPP Release 15~\cite{3GPP201}, extensive research and technical papers have been conducted to explore the feasibility and implications of incorporating \acrshort{ntn} into the standards. This scholarly endeavour examines various deployment scenarios and service types, playing a crucial role in facilitating the broader inclusion of HTS elements in 3GPP standards. With the introduction of 3GPP Release 17~\cite{3GPP2022}, this progress was further strengthened through significant steps taken toward the standardization of NTN. This involved developing specifications for system enhancements needed for NTN within the 5G infrastructure. The goal is to enable seamless, high-quality service delivery, particularly in regions where conventional terrestrial networks are unavailable. In the latest version of 3GPP Release 17~\cite{3gpp_tr_28808}, the main considerations for SatNet can be summarized as follows:
\begin{enumerate}
    \item Payload types: communication payloads on satellites can be transparent or regenerative, offering different capabilities such as signal conditioning, error correction, and network functions through OBP.
    \item (\acrshort{isl}): SatNet utilizes ISL to establish a dedicated mesh network, enabling connectivity and communication between satellites with regenerative payloads.
    \item Coverage enhancement: SatNet complements terrestrial networks, addressing limitations like physical constraints (maritime or aeronautical), economic reasons, or temporary unavailability of terrestrial networks (e.g., natural disasters or network overload).
    \item Content distribution: SatNet facilitates broadcasting, multicast (with content delivery acknowledgment), and unicast services, allowing the dissemination of content over large areas.
    \item Consideration of altitude and propagation delay: use cases for SatNets are affected by the actual altitude of satellite platforms and the associated propagation delay, which impact specific communication requirements and applications. 
\end{enumerate}

In conclusion, both 3GPP and DVB contribute significantly to the standardization of future SpaceNets. Their respective standards complement each other in providing comprehensive guidance on the design, implementation, and operation of SpaceNets, ensuring a high-quality and cost-effective ecosystem. 



\textit{Future Requirements:} While the 3GPP and other standards have contributed significantly to the advancement of mobile networks including SatNet, until now it does not specifically include provisions for HTS within its standardization efforts. That being said, the existing standardization may indirectly impact HTS systems through related aspects, such as the support for IP-based networking protocols and radio access technologies that could be utilized in conjunction with HTS deployments. However, for specific HTS-related standardization, other organizations, such as the International Telecommunication Union (ITU) may be more directly involved in developing standards and guidelines for HTS systems.  Therefore, for future regenerative satellite constellations with HTS, new standards, and novel designs and protocols are required to reach the goal of Terabits and beyond and to connect the unconnected.



\section{Patent Landscape in HTS Systems}
\label{sec:Patents}
Understanding the patent landscape related to SatNet systems is crucial, as they are key indicators of technological development and innovation in this field. They provide valuable perspectives on the speed and type of advancements, and potential future trends. By delving into the patents associated with HTS, we aim to highlight their significant role in the design of HTS systems.

Our exploration of the patent landscape related to satellite systems begins with routing methodologies in non-geostationary orbit satellites. The patent by Natarajan \textit{et al.}~\cite{natarajan2014method}  proposes a scheme for determining routes within an LEO satellite connecting a source ground user to one or more destination ground terminals. This scheme is designed to be employed by a ground-based network operations center, which handles user requests for establishing communication sessions between multiple ground terminals. These sessions are characterized by specific requirements such as bandwidth, priority, and duration. The scheme generates optimized network routes to facilitate the requested communication sessions, ensuring efficient and reliable connectivity. 
To determine routes for a data circuit between source and destination terminals using satellites in non-geostationary orbits, a new method was proposed in~\cite{kay2019radio}. In~\cite{kay2019radio}, routes are defined for data circuits by selecting satellites and establishing ISL based on time intervals and the shortest link duration. Both of these innovations focus on enhancing the routing capability and efficiency within SatNet.

Transitioning from routing methodologies in single-beam satellites, we delve into resource management in multi-beam satellite systems. In existing multi-beam satellite designs, resources are distributed in a fixed manner across different beams. This leads to inefficiencies and an inability to adapt to changing user service needs within each beam's coverage. On one end, surplus resources are wasted on light-load beams, while on the other end, heavy-load beams are under-resourced, failing to meet users' service requirements. This static resource allocation significantly lowers the resource utilization efficiency of the satellite resource manager. Therefore, it could be beneficial to implement a system design where frequency channels and resources are dynamically allocated to beams, driven by requests originating from the terminals~\cite{agarwal2013dynamic}. In~\cite{oh2014method}, data traffic is managed efficiently in a multi-beam satellite communication system by adaptively allocating resources based on each cell's required data transmission rate. Additionally, surplus resources from beams with remaining capacity are reallocated to beams requiring higher data transmission rates, optimizing the control of limited data traffic.
In~\cite{berman2000dynamic}, the authors use dynamic power allocation to maintain amplifier efficiency in a multi-beam satellite, catering to peak traffic demands while minimizing power consumption during low traffic periods. This is achieved by monitoring channel traffic and adjusting power accordingly. In one embodiment, the system conserves power and prevents amplifier saturation by adjusting the supply voltage based on the power level of the downlink signal, which reflects the channel's traffic. This power level sensed prior to amplification, guides the power supply to match input signal variations. Both patents highlight the need for dynamic resource allocation to adapt to changing user service needs within each beam's coverage.

 \textit{Future Requirements:} As previously outlined, an important challenge in HTS systems involves accurately routing traffic within the payload unit and balancing the load among several packet forwarders. However, when examining the current patent landscape, it is clear that the focus lies mainly on resource allocation among different beams and interference mitigation for HTS or single-beam satellite systems. Interestingly, to the best of our knowledge, there is a noticeable gap with hardly any patents addressing the specific issue of intra-payload routing and load balancing. This represents a significant opportunity for innovation in the field of HTS systems and underlines the need for greater focus on these crucial technical aspects.

\section{Multi-Channel Access Techniques in Satellites}
\label{multi-ch-access}

A primary operation executed by a satellite modem is multiplexing. In order to achieve high data transfer rates, a variety of bandwidth-sharing techniques are employed. These models determine how users share the satellite's resources effectively and can drastically impact the system's overall performance. These include FDMA, TDMA, code division multiple access (\acrshort{cdma}), spatial division multiple access (\acrshort{sdma}), orthogonal multiple access (\acrshort{oma}), NOMA~\cite{danishsatellite}, and RSMA.

The first generation of wireless standards employed basic TDMA and FDMA strategies. FDMA, although the easiest to implement, necessitates the assignment of unique frequency bands to each user. While TDMA is efficient, it requires framing and synchronization, which can pose issues in environments with substantial propagation delays. Additionally, these systems were often susceptible to eavesdropping by third parties~\cite{singh4g}. CDMA employs pseudorandom patterns to share the available frequency band in multi-user scenarios. Each user is given a unique random pattern, enabling multiple codes to be used simultaneously on the same frequency channel~\cite{danishsatellite}. Numerous studies have compared CDMA with FDMA and TDMA in satellite systems. While CDMA has proven to significantly boost capacity over FDMA and TDMA in terrestrial cellular networks, the same conclusion cannot be directly applied to LEO satellite applications due to the fundamental differences between terrestrial and satellite links~\cite{fu1999capacity}. Typically, CDMA is paired with a spread spectrum method, for instance, direct sequence (DS)-CDMA or frequency hopping (FH)-CDMA~\cite{oppermann1996coded,chan2007end,wei2017residual,khalife2018navigation}. \acrshort{mftdma}, also known as multi-carrier time division multiple access (MC-TDMA), is the prevalent technology system for contemporary broadband satellite communications. This hybrid multiple-access approach combines elements of both FDMA and TDMA systems. The MF-TDMA system adjusts the allocation of carriers and time slots based on the specific service requirements of the terminal, offering flexible support for various rate demands. The system partitions the channel into sub-carriers and employs TDMA to divide time slots within each sub-carrier. Consequently, each user is allocated a specific bandwidth within a defined time frame~\cite{yang2023research}. For multi-spot beam satellite systems that use the DVB-RCS2 return uplink, MF-TDMA has been demonstrated to enhance the overall throughput performance without introducing relevant processing overhead~\cite{bejarano2018mf}.  

Using SDMA, users are classified into several groups, where members of a group are simultaneously served by the satellite~\cite{chen2019user}. In multi-beam satellite systems, BF technology is commonly employed to execute SDMA, thereby enhancing the overall spectral efficiency of the system~\cite{ahmad2021zero}. However, to manage the growing number of satellite users within the confines of a limited spectrum, flexible resource assignment (RA) has been suggested as a solution. This approach, based on an OMA scheme, enables on-demand resource allocation to mitigate spectrum scarcity in satellite~\cite{miao2022research,abdu2021flexible}. In addition, cognitive radio (\acrshort{cr}) has been introduced to satellite systems to tackle the challenge of spectrum scarcity. This technology facilitates dynamic spectrum sharing between satellites in different orbits, such as allowing an LEO satellite to share the incumbent spectrum of a GEO satellite~\cite{tang2021resource}, or between satellite and terrestrial networks. Research has shown the performance improvements introduced by the CR strategy, and further studies have employed artificial intelligence for optimal spectrum resource management~\cite{liang2021realizing,pervez2021joint}. However, despite the improvements in spectrum utilization efficiency brought by RA and CR strategies, the potential of these improvements is restrained by the inherent limitations of the OMA scheme, which allows only one user access within a one-time slot or spectrum block~\cite{yan2023noma}.

\vspace{0.4cm}
\acrshort{noma} has been introduced as a potential successor to traditional OMA methods, aiming for greater spectral efficiency and extensive connectivity. In multi-beam satellite scenarios, the implementation of NOMA has revealed the significant advantages it offers compared to OMA~\cite{na2023multigateway,jeon2022advanced,ramirez2022contribution}. Essentially, NOMA enables various users to transmit over the same spectrum simultaneously~\cite{ali2022rate}. Given the challenges of supporting a vast number of Internet of Things (IoT) devices within limited spot beams, many studies have delved into the benefits and intricacies of implementing this in \acrshort{satcom}~\cite{zhu2019geographical,lu2020robust,chu2020robust}. The authors propose a combined strategy that integrates rate splitting (\acrshort{rs}) uplink NOMA with beamforming in the user link to enhance system spectral efficiency, using the available statistical channel state information (\acrshort{csi})~\cite{kong2022performance}. Through simulations, they examine the average throughput of the return link in an HTS system, comparing it to previous works~\cite{ahmad2020next,zedini2020performance, tegos2020outage,yan2019ergodic}. Their findings demonstrate the superior performance of the proposed scheme. For uplink communications, NOMA has the potential to mitigate the issues of frequency interference and coordination in multi-layer SatNet~\cite{ge2021joint}.  In the referenced work ~\cite{liu2019qos}, the authors suggested the use of a multi-beam satellite for the industrial Internet of Things (IIoT), with the goal of ensuring spectrum access and enabling interconnections over vast areas, particularly for nodes situated in remote regions. They employed NOMA within each satellite beam to enhance the IIoT's transmission rate given the constraints of limited spectrum resources. Their findings demonstrated that the aggregate transmission rate of NOMA was significantly higher than that of OMA. Moreover, they noted that more power was allocated to the node with the weaker channel to ensure its transmission performance amidst high levels of interference. Recently, grant-free NOMA became a promising multiple access technique that leverages the advantages of random access (RA) and NOMA~\cite{shahab2020grant}. In grant-free NOMA, several users engage in uplink instantaneous transmissions without the need for a specific grant, resulting in a notable reduction in signaling overhead~\cite{9022993}. On one hand, this approach can be effectively applied to satellite-based IoT scenarios, particularly in situations where minimizing the overhead associated with controllable interference and signaling interactions is crucial~\cite{ye2022help}. 

Initially, RSMA was explored for application in terrestrial networks~\cite{dai2016rate}, but it has recently been extended to Satcom systems, in particular to multi-beam satellites~\cite{vazquez2018rate, kong2022combined, schroder2023comparison,kong2021beamforming}, and to satellite-aerial integrated networks~\cite{yin2022rate}. RSMA, which uses linear RS at the transmitter and the successive interference cancellation (SIC) technique at the receiver, has been recognized as a novel and overarching framework for designing and optimizing wireless transmission~\cite{kong2022combined}. In RSMA, every user's message is divided into a private and a common segment. The common segments of all users' messages are jointly encoded into a common data stream for all users. Depending on the degree of correlation between the user channels and imperfection errors, varying amounts of power are allocated to the common and private segments~\cite{schroder2023comparison}. Specifically, RSMA is a technique that partially interprets interference as noise while partially extracting information from it~\cite{cui2023energy}. This approach forms a bridge between SDMA and NOMA~\cite{mao2018rate}. The authors in~\cite{kong2022combined} present a resilient uplink transmission strategy that integrates BF with RSMA to accomplish a high throughput in a multi-beam satellite system, under the assumption that only imperfect CSI is accessible. The goal is to reduce the total transmission power such that each user's individual transmission power budget and QoS requirements are satisfied. Recently, the authors in~\cite{liu2023energy} investigated the issue of maximizing energy efficiency (EE) with RSMA in multi-beam satellites, considering a more realistic and generalized imperfect CSI at the transmitter (CSIT). The simulation results display that RSMA offers superior performance compared to SDMA under both perfect and imperfect CSIT conditions.

\textit{Future Requirements:} To conclude, RSMA holds significant potential in realizing the requirements of advanced SatNet including higher throughput, optimal power consumption, low latency, and hybrid systems by boosting the adaptive use and efficiency of the available bandwidth. In addition, grant-free NOMA provides high performance in LEO constellation-enabled IoT devices~\cite{zhou2021joint}. However, the latter technique has not been studied for HTS with multiple beams. Therefore, we expect that non-orthogonal techniques especially RSMA and grant-free NOMA will be the leading candidates in future regenerative constellations with HTS to achieve multi-connectivity with high data rates up to Terabits.

\section{Network Routing and Scheduling in HTS Systems}
\label{sec:Routingandsch}
\subsection{Network Routing}
\label{sec:Routing}
The problem of point-to-point routing remains a challenging issue in SatNets that comprise a large number of satellites. The unique characteristics of satellites, including limited onboard processing and storage capabilities, frequent and abrupt changes in network topology, uneven distribution of data flow, a high bit error rate, and prolonged delays in ISL, pose significant difficulties for efficient routing. As such, devising effective routing solutions for SatNets is of great importance in realizing their potential for various applications~\cite{xiaogang2016survey}.

In the current literature, routing has been extensively studied from the perspective of both single-layer and multi-layer satellite constellations, focusing mainly on single-beam satellite systems~\cite{8377540,li2015routing,he2022cross,xu2022link,yin2023joint, madoery2023routing}. However, there is a notable gap in the literature when it comes to routing in multi-beam satellite systems. Despite the increasing interest and deployment of such systems, there is a dearth of research on the routing protocols and algorithms that can efficiently manage the traffic flows in these systems. This gap in the literature presents an opportunity for researchers to explore and develop novel routing solutions that can optimize the performance of single multi-beam satellite systems and can contribute to the advancement of SatNet technologies. 


Recently, telecommunication industrial companies including SpaceX, Telesat, and OneWeb have shown a growing interest in LEO satellite constellations due to their worldwide connectivity benefits and the scarcity of orbital resources~\cite{pachler2021updated}. For example, Starlink is planning to launch 12,000 with a possible extension to 42,000 satellites~\cite{zheng2022leo} and Oneweb is planning a network of 643 satellites with 30 satellites planned for resiliency and redundancy~\cite{oneweb2023}.
The satellite constellation is equipped with four ISLs, facilitating communication both within and across planes. Using advanced electronically steerable antennas, such as phased-array antennas, the ISL reassignment method can be easily implemented~\cite{cowley2010phased}. These constellations rely on RF links for both feeder and user connections, implying that the restriction of bandwidth could pose a challenge in improving the data rate~\cite{calvo2019optical}. Hence, despite the time-consuming establishment process, optical ISL have been incorporated into LEO constellations, owing to their ability to support large capacity transmissions ranging from tens to hundreds of Gbps~\cite{smutny20095,kaushal2016optical,carrizo2020optical,10032696,10119016}.


Given the huge topology of planned constellations, traditional static algorithms struggle to compute the optimal routing strategy. Therefore, broadcasting the real-time traffic is critical. Furthermore, as data traffic increases due to a variety of applications, each with their unique QoS demands, the network performance is anticipated to confront issues such as congestion and overload. Additionally, different from terrestrial networks where one can often boost network capacity by upgrading congested links, this approach is not feasible in SatNets~\cite{9709357}. Hence, it becomes essential for designers of satellite constellations to devise proficient routing algorithms to allocate and optimize the usage of available resources effectively. In the current literature, various studies have proposed different routing strategies in satellite constellations with ISL~\cite{zheng2022leo,wang2023inter,razmi2023board, dong2022load,yang2023interruption}. In order to satisfy the QoS demands of various communication services, a QoS-aware routing algorithm (QoSRA) was proposed in~\cite{zhang2021qosra}. This algorithm is based on a traffic scheduling framework that merges the priority queue (PQ) and weighted round-robin (WRR) approaches. The algorithm not only meets the QoS needs of high-priority services, but it also prevents low-priority services from being overlooked due to their less competitive nature. Additionally, the algorithm utilizes the capabilities of the software-defined SatNet model to manage bandwidth allocation effectively. Simulation comparisons show that QoSRA can guarantee the QoS requirements for various service types, and it outperforms other routing algorithms in terms of average end-to-end delay, packet loss rate, and data throughput.


In the context of routing, the conventional approach of shortest path routing aims to minimize the number of hops between source and destination nodes. However, this strategy can lead to congestion in certain parts of the network, resulting in low overall throughput. To strike a balance, a combination of the two routing policies was proposed in~\cite{gounder1999routing}. By dedicating a portion of the bandwidth of each ISL to real-time data and allowing other data types to share the remaining bandwidth, real-time data can be routed using the shortest path technique while other data types can follow the multi-commodity flow approach. This balanced approach can help achieve optimal results in terms of both latency and throughput. To summarize, traditional routing algorithms often neglect important factors such as congestion and network capacity limitations. Convex optimization techniques offer a powerful toolset for modelling and solving complex optimization problems with provable guarantees~\cite{boyd1977convex}. Combining routing algorithms with optimization can enhance the efficiency and effectiveness of traffic assignment algorithms. An example of such integration is the application of the Frank-Wolfe algorithm. The Frank-Wolfe algorithm, also known as the conditional gradient method, is an iterative optimization technique that solves convex optimization problems. By leveraging the principles of the Frank-Wolfe algorithm, route choices, and traffic flows can be optimized by iteratively updating the traffic assignment based on the gradient of the objective function~\cite{chen2001effects, lee2001accelerating}. This approach allows for the effective assignment of traffic in the network while considering network constraints, leading to improved network performance and resource utilization.

To address the requirements of diverse service transmissions within energy-constrained SatNets, several research papers propose energy-aware traffic classification routing algorithms that leverage traffic characteristics. This approach enables the optimization of routing decisions to effectively manage limited energy resources while accommodating the specific demands and priorities of different types of services. By considering the varying energy levels, the proposed scheme in~\cite{hao2020satellite} optimizes routing strategies, resulting in enhanced data transmission for reliability services. 
In~\cite{7572177}, Yang \textit{et al.}  propose an energy-efficient routing algorithm for SatNets. The authors address the challenge of high energy consumption in SatNets by designing routing algorithms that consider both the energy cost and the QoS requirements of the network. The proposed algorithm considers the energy cost of each link and calculates the optimal path for data transmission based on the network topology and QoS requirements. The authors also propose a hybrid routing scheme that combines both the shortest path and minimum energy routing to further reduce energy consumption. The effectiveness of the proposed algorithm is evaluated through simulations, which show that it achieves a significant reduction in energy consumption while satisfying the QoS requirements of the network.

\textit{Future Requirements:} Routing in HTS satellites is a complex process that involves directing data packets between different ground terminals via the SatNet. Unlike traditional satellites, which use \acrshort{fdma} or \acrshort{tdma} techniques to allocate bandwidth to specific users, HTS systems use a combination of frequency reuse, spot beams, and dynamic resource allocation to provide high throughput and efficient utilization of satellite resources.
The success of HTS systems is heavily dependent on the efficient routing of traffic across the network. In this context, it is expected that existing routing algorithms developed for terrestrial networks can be leveraged for HTS systems as well. This is due to the similarity between modem banks and routers or asynchronous transfer mode (ATM) switches, as the payload consists of multiple packet forwarder~\cite{baudet2017innovative} or multiple modem banks that can be controlled using \acrshort{sdn} techniques. In such a system, the routing process will take place within the payload to find the best path between all beams or to send data to another satellite. The routing algorithms developed for terrestrial networks can be modified and optimized to account for the unique characteristics of HTS systems, such as high latency and limited bandwidth, and they provide a strong starting point for developing effective routing solutions for SpaceNets.

\subsection{Scheduling }
\label{sec:Scheduling}

In addition to packet forwarding and routing, network devices must perform extra functions to provide network-level QoS. Classifications, queuing and scheduling, policing, and buffer shaping and management are some of these features~\cite{el2003evolution}.

Addressing multiple users in the same frame raises precoding complexity as multiple spatial signatures must be assumed for a single codeword. Users within the same beam will have separate channel vectors, resulting in poor precoding efficiency~\cite{7811843}. Therefore, novel techniques for joint user scheduling and precoding in multi-beam satellite systems have been addressed in the current literature~\cite{8401547,arapoglou2016dvb,zhang2021joint,honnaiah2021demand,kibria2019precoded}. It is worth noting that the combination of scheduling and precoding yields better performance. In~\cite{el2003evolution}, the authors presented and designed a geographical scheduling algorithm (GSA) for multicast precoding in multi-beam HTS. Based on the idea of scheduling users in similar locations in their associated beams, this algorithm improves multicast and unicast precoding performance leading to a significant overall gain in the system. Furthermore, beam scheduling in the context of BH involves designing an optimal beam illumination pattern. From a higher-layer perspective, scheduling can be also integrated with network slicing~\cite{kisseleff2020radio}.

To efficiently allocate bandwidth among various users, dynamic bandwidth allocation (DBA) in HTS provides broadband communication services using multiple spot beams, which enables the delivery of high data rates to users on the ground. The primary goal of DBA is to ensure optimal utilization of satellite resources and improve the QoS for users~\cite{8372935}. However, the increased flexibility of HTS systems also leads to a higher level of complexity in dynamic resource allocation (DRA).
To overcome the limitations of the decision-making optimization problem in \acrshort{dra} and beam scheduling in HTS systems, a deep reinforcement learning (\acrshort{drl})-based approach is adopted~\cite{huang2022sequential}. This approach addresses the slow convergence time and meets the real-time requirements of modern applications. In addition, it addresses signal transmission latency, data packet loss ratio, and power energy consumption load by formulating a multi-objective optimization problem. Simulation results confirm the effectiveness of the proposed algorithm in addressing various user traffic demands and demonstrate its superiority over existing methods.
Lately, deep learning (DL) has emerged as a promising approach for complex resource allocation in BH systems. \acrshort{dl} techniques have been successfully applied in this context to address resource allocation challenges~\cite{lei2020deep, lin2022dynamic, hu2020dynamic}. The authors in~\cite{lei2020deep} were the pioneers in conducting initial investigations on applying DL to optimize BH systems. In this study, they examine a combined learning-and-optimization scheme to offer a fast, feasible, and nearly optimal solution for scheduling in BH systems. The authors in~\cite{lin2022dynamic} propose a DRL-based scheme for dynamic beam pattern and bandwidth allocation. A cooperative multi-agent DRL (MADRL) framework is utilized to handle the joint allocation problem, where each agent is responsible for a specific beam. The agents collaborate to maximize throughput and minimize delay fairness. Simulation results show that the offline-trained MADRL model achieves real-time allocation and exhibits good generalization under increasing traffic demand. In~\cite{hu2020dynamic}, the optimal policy for BH in a DVB-S2X satellite was investigated, aiming to ensure fairness, minimize transmission delays for real-time services, and maximize throughput for non-instant services. A novel multi-action selection method based on double-loop learning (DLL) is proposed to address the problem of high-dimensional actions. The proposed method intelligently allocates resources based on user requirements and channel conditions, as demonstrated by realistic evaluation results. In summary, the application of DRL and DL techniques in DRA and BH systems demonstrates their potential to enhance resource allocation efficiency, optimize performance metrics, and provide adaptive solutions for modern satellite communication networks.

Efficient scheduling in multi-beam satellite systems is vital for maximizing system capacity, ensuring reliable communication services, and optimizing resource utilization. Advanced algorithms and optimization techniques are utilized to achieve these goals, enabling the satellite system to operate efficiently and effectively. These algorithms assign users or service areas to specific beams based on their location and coverage requirements. They also dynamically adapt the resource allocation and beam assignments in response to changing conditions, such as user demand and system congestion. By leveraging these advanced techniques, efficient scheduling ensures that multi-beam satellite systems operate at their maximum capacity, provide reliable communication services, and effectively utilize available resources.

\textit{Future Requirements:} In the current literature scheduling resource allocation between multiple beams for HTS has been extensively studied. However, as future payloads of HTS are expected to be composed of multiple modem banks, scheduling and efficient resource management within the payload are needed. Therefore, there is a gap in possible scheduling techniques within the satellite OBP. Finally, we anticipate that integrating traditional scheduling algorithms with artificial intelligence (AI) and SDN techniques will enable the realization of global connectivity at Terabit speeds, facilitating dynamic and adaptable resource allocation within the satellite's payload.

\section{Load Balancing and Quality of Service}
\label{sec:LoadBalancing_QOS}
\subsection{Load Balancing}
\subsubsection{Importance of load balancing in HTS}

\label{sec:LoadBalancing}

The primary aim of load balancing is to maximize the use of existing network capabilities to decrease the likelihood of traffic bottlenecks. This ideally results in reduced delays and packet loss. However, if alternative paths are not selected wisely, it could potentially result in increased propagation delays~\cite{lee2002survey}. Therefore, load balancing can be a critical factor in the performance and efficiency of HTS systems. Typically, HTS systems use multiple spot beams to provide coverage over a wide area, and traffic demands across these beams can vary significantly. Load balancing helps to ensure that the available resources are utilized optimally and evenly across the entire coverage area, which is essential for providing reliable and high-quality service to all users.

\vspace{0.4cm}

In an HTS system, load balancing can be accomplished through various methods such as adaptive coding and modulation, beam shaping, and traffic management algorithms. These techniques aim to enhance the utilization of available resources like spectrum and power, while also maintaining an equitable distribution of traffic among spot beams. This becomes increasingly significant in high-traffic-demand systems, where system performance may be adversely affected by congestion and interference.

The distribution of traffic across the spot beams must be carefully managed to ensure that the system is scalable, flexible, and resilient. This requires the use of sophisticated network management and control systems that can dynamically adjust the routing and distribution of traffic based on real-time network conditions and user demand.

\subsubsection{Load Balancing in HTS Using Machine Learning}
\label{sub:LBwithML}

HTS systems are increasingly employing machine learning (\acrshort{ml}) algorithms to optimize their performance in terms of load balancing, leveraging ML's ability to analyze large volumes of historical data and predict future patterns~\cite{gures22}. As a subset of artificial intelligence, ML is proficient in predicting future trends based on historical data analysis, which is crucial in preempting network congestion and proactively reallocating resources~\cite{Liu2021}.

In particular, reinforcement learning (\acrshort{rl}), a type of ML, stands out in its efficacy in load balancing. RL operates on the principle of rewarding or \say{reinforcing} actions leading to the desired outcome, training the system to make optimal decisions. The RL model is aptly suited to HTS, where the objective is to determine the most effective strategy for resource allocation based on past performance data~\cite{peiliang22}. The model tweaks the resource distribution across different beams and monitors the impact on system performance with a specific aim, such as maximizing system throughput~\cite{jingxu2023}.

In a study by Liu \textit{et al.}, the authors introduced a routing algorithm based on deep deterministic policy gradient (DDPG), an RL algorithm for low-orbit SatNets. The results revealed the DDPG-based algorithm surpassed traditional algorithms in terms of throughput, packet loss rate, average delay, and load distribution index~\cite{Liu2021}.  Similarly, Dai \textit{et al.}  proposed a DRL-based power allocation model for HTS, highlighting its capability to cope with unpredictable dynamic channel conditions~\cite{9625395}. The model leveraged the model-free nature of RL to transform these conditions into a power allocation problem in the HTS system. Zuo \textit{et al.}  further expanded this idea by introducing a decentralized load balancing routing algorithm for Low Earth Orbit SatNets (LEO-SN) based on DRL. The model factored in the one-hop satellite node's queuing delay, storage space, communication bandwidth, and propagation delay. The proposed algorithm demonstrated faster convergence and improved performance in terms of packet loss rate and transmission latency compared to conventional methods~\cite{peiliang22}. Xu \textit{et al.}  developed a novel DRL architecture for dynamic power and bandwidth allocation in multi-beam satellites. The proposed method optimized unmet system capacity demonstrating superior performance and significantly reducing computation time compared to existing methods~\cite{jingxu2023}.

The application of these ML algorithms is driving the move towards more intelligent and autonomous satellite systems with the ability to dynamically adapt to changing conditions and demands, significantly improving their efficiency in load balancing. As a result, these advancements in ML-based load balancing methods are revolutionizing the field of satellite communication, enabling more robust and adaptive communication systems.

\subsubsection{The Potential of Queuing Theory in Load Balancing in HTS}
\label{sub:LBwithQT}

The high demand and the ever-increasing traffic in SatNet, highlight the importance of efficient and reliable packet scheduling~\cite{zeephongsekul2006}. One area that has been under-researched is the application of queuing theory in load balancing within HTS. Although a vast body of literature exists on using queuing theory to model terrestrial networks~\cite{zeephongsekul2006, taixin2017}, to the best of our knowledge, there is a gap in applying these theories to HTS systems, which would potentially improve their Service Level Agreements (SLAs).

Incorporating queuing theory into HTS systems can provide a mathematical model to forecast traffic. This thereby enables proactive resource allocation, thus improving the overall performance of the system. Predicting periods of high demand would prevent bottlenecks and preemptively alleviate network congestion~\cite{9745153}. In particular, priority queuing plays a critical role in ensuring \acrshort{qos} by classifying and prioritizing different types of traffic based on their requirements. Priority queuing can also significantly improve the SLA of an HTS by ensuring that high-priority traffic is served first~\cite{lemeshko2023}.

Lemeshko \textit{et al.} ~\cite{lemeshko2023} discussed a two-level hierarchical queue management method based on priority and balancing. The study further emphasized the role of effective queue management in mitigating congestion and ensuring balanced and priority-based packet flow distribution. Another important aspect to consider is finite buffer queuing, particularly when addressing delay performance in LEO satellite systems~\cite{marcano22}. Hernández Marcano \textit{et al.} 's work demonstrated how short buffers can ensure less than 5-10\% packet loss with tolerable delays, highlighting the importance of a novel approach to queuing in LEO satellite systems~\cite{marcano22}. Moreover, novel approaches proposed~\cite{Chan_Su_2002} to minimize queuing delay by considering transmission constraints can be beneficial in the SatNet system context. Their method suggests configuring multiple queues and using a search order table to schedule the transmission of packets based on system constraints. 

\subsubsection{Load Balancing in HTS Using Optimization Algorithms}  
\label{sub:LBwithOPT}
  Metaheuristic optimization algorithms have been effectively employed in many load balancing applications, especially in 5G networks, with promising results. For instance, in~\cite{krishnamoorthy22}, a customized genetic algorithm was proposed for load balancing in 5G ultra-dense networks, showcasing the effectiveness of metaheuristics in such scenarios. However, the application of these methods is not limited to 5G technologies. The emergence of HTS provides a compelling case for the application of these techniques in the satellite communication domain. Currently, various studies have highlighted the advantages of employing optimization algorithms for load balancing in HTS. The authors in~\cite{yuanzhi2022} proposed a novel heuristic algorithm for optimizing resource allocation in multi-beam SatNets, which effectively improved network efficiency and reduced beam interference. Additionally,~\cite{pachler22} introduced a metaheuristic optimization algorithm based on Particle Swarm Optimization (PSO) to solve the complex resource allocation problem in multi-beam HTS, demonstrating its superiority over traditional methods.

\textit{Future Requirements:} To summarize, effective load balancing also enables scalability, flexibility, and resiliency in the network architecture, which is essential for meeting the evolving needs of the SpaceNets industry. Despite these advancements, there is no work that has explicitly considered optimization algorithms for load balancing within an HTS, particularly among multiple modem banks to avoid congestion and bottleneck. This area presents a unique and largely unexplored opportunity. Given the success of metaheuristics in analogous applications, convex optimization, and online optimization in different areas, the same success can be projected onto the HTS domain.

\subsection{Quality of Service}
\label{sec:QoS}

The \acrshort{qos} of an HTS system is a measure of its ability to meet user expectations for service availability, reliability, and performance. These expectations can vary widely depending on the application, the user's location, and the service provider's requirements. For example, a video streaming service may require low latency and high bandwidth, while a remote healthcare application may require high reliability and availability. To ensure that the QoS of an HTS system meets user expectations, a number of factors must be considered. These include the design of the satellite system, the availability of spectrum resources, the network architecture, the capacity management, control mechanisms, and the implementation of QoS policies.

Existing literature related to precoded HTS includes, specifically addressing parameters like minimum signal-to-noise ratio (SNR) or minimum throughput per user as given in~\cite{qi2019energy,zhang2020joint}. Nevertheless, the challenge arises when dealing with uneven QoS demands, especially in scenarios with high QoS requirements and limited satellite resources. The authors in~\cite{zheng2012generic} proposed a novel method for preserving each user's QoS requirements by formulating a new precoding design in a multi-beam satellite. They employed an alternating optimization algorithm, which led to an improvement in data throughput via the proposed iterative process. However, the solution's scalability is limited. Specifically, a max-min fairness optimization framework fails to provide an adequate QoS level for a large-scale system with a high number of users. 
In a complex system accommodating a considerable number of users with varied Quality of Service QoS requirements, there is a high likelihood that at least one user will experience severely unfavorable channel conditions, or their QoS demands will exceed what the limited radio resources can provide. As a result, the existing solutions become untenable, leading to congestion and subsequently an infeasible problem. This scenario underlines the challenges in maintaining QoS across a wide user base in resource-constrained environments. To address this issue, two methodologies are proposed: (1) a model-based approach, and (2) a data-driven approach. On one hand, model-based methods are generally recognized for delivering precise solutions. On the other hand, data-driven approaches, inspired by advancements in ML, have demonstrated faster convergence towards near-optimal solutions~\cite{van2020power, sun2018learning}. In~\cite{bui2022robust}, the authors address the congestion problem in demand-based optimization for multi-beam, multi-user satellite communications. Formulating a multi-objective optimization, the authors maximized the sum rate and the number of satisfied users across all channel conditions, while sharing the same time and frequency resource plane. Prioritizing QoS requirements, they designed heuristic algorithms that can effectively resolve the optimization problem within both feasible and infeasible domains, under power limitations and individual QoS demands. Leveraging the water-filling method and linear precoding technique, the proposed framework significantly boosts the number of satisfied users compared to some benchmarks. Additionally, the deployment of a neural network drastically reduces runtime, making real-time power allocation and user satisfaction control feasible in satellite systems that require millisecond-scale updates due to variations in user scheduling or individual demands.


The load balancing strategies and QoS maintenance techniques significantly impact the overall performance and efficiency of the HTS systems. An effective load balancing mechanism can optimize the use of system resources, ensuring uniformity in service provision. Simultaneously, it helps prevent network congestion, reduces the likelihood of service degradation during peak demand times, and increases system resiliency. Load balancing directly influences the scalability of an HTS system~\cite{kansal2012}. As user demand grows or fluctuates, a well-balanced system can adapt without overloading individual spot beams or facing significant performance degradation. This capacity to handle increased demand while maintaining performance, also called elasticity, is a key factor in the overall efficiency of an HTS system.

QoS policies also profoundly affect HTS performance and efficiency. They ensure that the system can deliver the required service levels to meet the diverse needs of different applications and users. A well-implemented QoS policy can prioritize critical traffic, guarantee service availability, and manage network resources in a way that maximizes system efficiency~\cite{https://doi.org/10.1002/sat.765}. Furthermore, QoS mechanisms play a crucial role in maintaining user satisfaction~\cite{7463014}. The authors of~\cite{van2021user} introduced an innovative user scheduling design that ensures individual QoS requirements are met while maximizing system throughput. This is accomplished by using precoding to minimize mutual interference. However, the combinatorial optimization structure imposes a significantly high computational cost to reach the global optimum, even with a limited number of users. As a result, they suggest a heuristic algorithm that provides an efficient local solution and manageable computational complexity, making it suitable for large-scale networks. Numerical results confirm the effectiveness of their proposed algorithm in scheduling numerous users, achieving superior sum throughput compared to other benchmarks. Additionally, the QoS requirements for all users are effectively met.

Finally, with the aim of direct satellite-to-device connectivity, it is expected that satellite systems to have the potential to establish direct connectivity with billions of mobile user equipment~\cite{tuzi2023satellite}. Currently, industrial companies are actively pursuing direct connectivity solutions, providing essential emergency services. These initiatives involve leveraging existing satellite constellations, exemplified by Apple~\cite{apple_globalstar} and Huawei~\cite{huawei_satellite_texting}, or exploring and testing novel approaches, as seen with Lynk~\cite{lynk_cell_tower}. Therefore, this immense potential raises concerns about potential congestion and bottlenecks. In this context, QoS and traffic prioritization will play a pivotal role as key enablers for future SpaceNets.

\textit{Future Requirements:} Upon reviewing the existing literature pertaining to HTS systems, it becomes clear that there is a significant lack of emphasis on two critical aspects: QoS and priority management. Both of these elements are essential for ensuring the reliable and efficient functionality of HTS systems. Therefore, considering traffic prioritization within the satellite payload is essential for future space systems. We also expect that combining load balancing and traffic prioritization by using the WRR algorithm~\cite{katangur2022priority} to schedule tasks will optimize the network efficiency by enhancing both throughput and the performance of individual modem banks.


\section{Software Defined Networking}
\label{sec:SDN}

In what follows, we provide an overview of existing research that incorporates SDN into SatCom, with a particular emphasis on HTS systems. These studies serve as foundational references that can be adapted and extended to suit the unique requirements of future regenerative satellite constellations.

In the past, network devices, such as switches and routers, housed both control and data forwarding functions, introducing challenges in large-scale network management~\cite{mckeown2008openflow}. The advent of SDN sought to remedy this by centralizing control, thereby decoupling the control and data planes~\cite{6994333}. This innovation expanded the horizons of network infrastructures, extending its influence from terrestrial domains to the vast expanse of satellite systems~\cite{7060482}.

Lately, SDN has emerged as a promising approach to enhance the efficiency of satellite systems~\cite{jiang2023software, rossi2015future}. Unlike traditional solutions, the SDN-based approach offers distinct benefits. The principle of separating the control and data planes reduces the load on individual satellite nodes. In the data plane, satellites handle straightforward tasks like data forwarding and network configuration, which require minimal computational and communication resources. Meanwhile, the more complex responsibilities, such as resource allocation, routing strategy design, and network management, are offloaded to the SDN controller. This controller, usually stationed in a more robust satellite or ground station, is better equipped to handle these intricate tasks. This division of tasks optimizes the use of satellite resources and potentially increases the network's overall performance~\cite{jiang2023software}.


The utilization of SDN in HTS management promises significant improvements in bandwidth usage, network optimization, and operational cost reduction. By shifting the control logic from the hardware to the software layer, SDN provides a flexible and programmable networking environment~\cite{8258968}. This decoupling empowers operators with better control over their HTS networks, enabling dynamic network configuration and efficient resource allocation. An example of this is the development of SDN-based BH in HTS, which allows for flexible and efficient bandwidth distribution based on user demand.

In the world of SDN, latency is a challenge that is amplified when considering HTS systems. Traditional terrestrial SDN deployments benefit from negligible packet travel times between the switch and the controller. In contrast, satellite systems, especially those using GEO satellites, face significant transmission delays due to the vast distances involved~\cite{8473417}.

Moreover, when the SDN controller is stationed on the ground, an additional layer of latency is introduced. The time it takes for signals to travel between the satellite and Earth becomes a non-trivial factor~\cite{8286925}. This combination of transmission times, paired with the SDN model's centralized decision-making, heightens the latency challenges. Consequently, future SDN solutions for HTS must focus on optimizing controller placements or strategizing to manage tasks at the satellite level without constant recourse to a central controller.



Several technologies are fundamental in the integration of SDN within HTS systems. OpenFlow, a well-known SDN standard, enables the communication between the controller and the forwarding plane, providing a foundation for network programmability~\cite{10.1145/1355734.1355746},\cite{xu2019openflow}. OpenFlow-based satellites can dynamically manage traffic flows and reduce network latency. 

Moreover, \acrshort{nfv} complements SDN in HTS by abstracting network functions from dedicated hardware~\cite{8060513}. This abstraction results in more versatile and cost-efficient networks. NFV has been adopted in HTS for virtualizing network functions such as routing and firewalls, paving the way for cloud-based SatNets~\cite{FERRUS201695}. The authors in~\cite{bertaux2015software} delved into the benefits of incorporating network programmability and virtualization using SDN and/or NFV. They examined their impacts on a typical satellite system architecture through the lens of four distinct use cases. The authors of~\cite{bao2014opensan} proposed a SatNet architecture built on the concept of decoupling the data and control planes, an approach that yields improved efficiency, refined control, and greater flexibility. Subsequent works have extensively explored the advantages and technical challenges associated with integrating SDN/NFV into SatNets and beyond~\cite{gardikis2017towards,li2016using,barakabitze2022sdn}. These comprehensive studies provide a range of use cases, opportunities, scenarios, and research challenges. Importantly, they identify SDN as a promising catalyst for the evolution of service delivery over integrated satellite-terrestrial networks~\cite{akyildiz2015softair}.


As SDN technologies continue to evolve, we can expect significant advancements in HTS. The integration of ML algorithms with SDN controllers, for example, may allow for predictive network management~\cite{8444669},~\cite{9493245}. This could enable proactive adjustments in HTS networks based on forecasted data traffic patterns, enhancing network reliability and efficiency. Furthermore, the advent of quantum SDN could offer unprecedented levels of security and speed in satellite communications~\cite{aguado2016quantum}.


\begin{table*}
\caption{From regular satellite to EHTS.}
\centering
\begin{tabular}{| p{3cm} | p{4cm} | p{4cm} | p{4cm} |} 
\hline
\textbf{Feature} &  \textbf{Single-Beam Satellite } & \textbf{HTS} & \textbf{EHTS} \\
\hline 
Frequency & C-band, X-band, ku-band, ka-band &ku-band, ka-band& ku-band, ka-band \\
\hline
Data rate &1-10 Gbps per satellite  & Up to 300 Gbps depending on size and frequency. Up to 2 Gbps  within a spot beam & Extremely High (Terabits) \\
\hline
Global coverage & Wide coverage & remote areas, sea and air & Global coverage: space-air-ground-sea \\
\hline
Number of beams & Large single-beam &  Multiple narrow beams (limited) & Multiple narrow beams (significant increase) \\
\hline
Multiple Access Technology & FDMA, TDMA & MF-TDMA & Advanced schemes: NoMA \& RSMA\\
\hline
Deployment &Early space communications, TV broadcasts & Broadband services, remote area connectivity, dedicated for mobility & Future broadband services, universal connectivity, dedicated for mobility and regenerative constellations\\
\hline
\end{tabular}
\label{tab:eHTS}
\end{table*}
The incorporation of SDN in HTS holds the potential to transform satellite communications. Enhanced network programmability will allow operators to rapidly respond to changing conditions and customer needs, resulting in higher customer satisfaction and more efficient resource usage~\cite{7302778}. Moreover, reduced operational costs and increased scalability afforded by SDN will enable the expansion of satellite services to broader markets, including rural and underserved areas.

As the fusion of SDN and HTS systems continues to gain momentum, there are several promising avenues for exploration. Firstly, the potential integration of edge computing within SatNets can further capitalize on the advantages offered by SDN~\cite{7956189}. Secondly, there is a burgeoning interest in incorporating blockchain technology within SDN-managed SatNets. Such integration could provide enhanced security, transparent transactions, and a decentralized control mechanism, enhancing trust and reliability in the network~\cite{8030491}. These directions, among others, signify the evolving landscape of satellite communication, with SDN playing an increasingly pivotal role in shaping the future.

However, these benefits will not come without challenges. The integration of SDN into HTS systems will require significant investments in infrastructure, training, and research. Security risks may also increase due to the centralized nature of SDN, necessitating advanced security measures~\cite{NISAR2020100289}. 

\textit{Future Requirements:} To achieve the objective of autonomous resource management and facilitate dynamic and adaptive resource allocation within future SpaceNets, we foresee the significance of developing a new SDN architecture. This architecture will play a pivotal role in monitoring and managing traffic within the satellite, particularly among different modem banks.



\section{EHTS: A Vision for the Future of Satellite Systems}
 \label{sec:EHTS}

As our demand for data and connectivity grows and satellite technology continues to advance, we anticipate a need for even more sophisticated satellite systems, evolving towards EHTS. These next-generation satellites are expected to provide unprecedented terabits-level data rates. The envisioned EHTS would employ even narrower beam technologies, allowing for multiple beams to illuminate a single geographical zone as shown in Figure \ref{figure_ehts}. In addition, using advanced cognitive radio techniques, multiple satellite systems could cover simultaneously the same area depending on the traffic distribution. This ensures that high-throughput benefits extend to all users, even those located at the beam edges. Moreover, to optimize spectral efficiency, EHTS are anticipated to leverage the use of non-orthogonal multiple access techniques including grant-free NOMA and \acrshort{rsma}. These advanced multiple-access techniques aim to maximize the efficient use of available resources, thus contributing to the ultimate goal of global connectivity. The envisioned architecture for EHTS and their main features are summarized in Figure \ref{figure_ehts} and Table \ref{tab:eHTS}.

\begin{figure*}
  \centering
     \includegraphics[width=1\textwidth, height=12in,keepaspectratio]{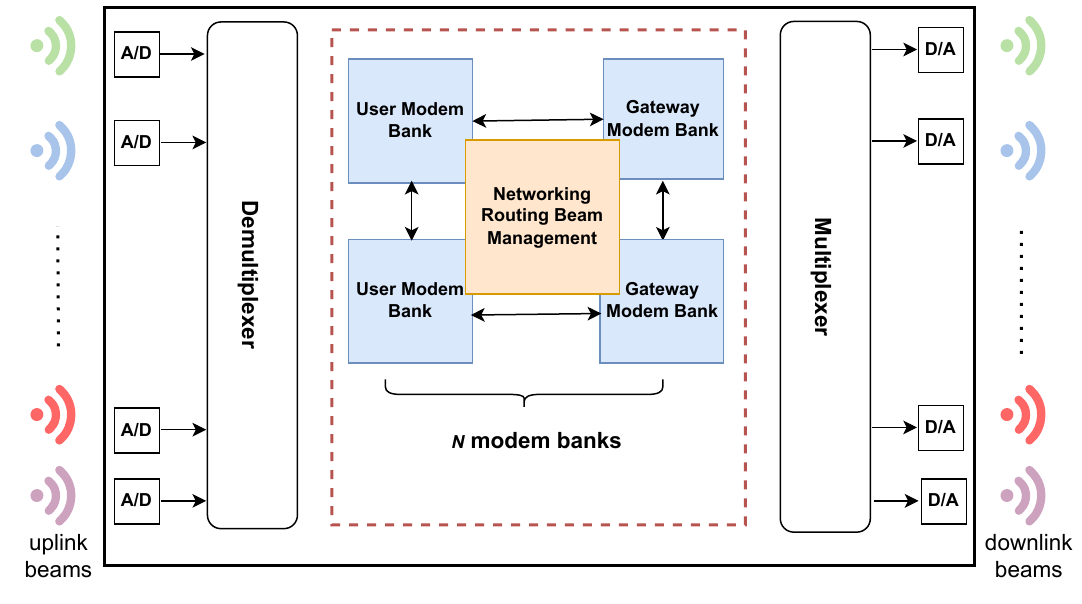}
   \caption{Satellite Communications Payload Overview.}
  \label{figureOBP}
\end{figure*}

In addition, based on the payload overview presented by Kuiper~\cite{hindin2019technical}, we propose a novel architecture for future satellite payloads as depicted in Figure \ref{figureOBP}. In this architecture, we consider different types of modems and a packet routing and beam management sub-system. These modems will support modulation, adaptive coding, and error correction. In addition, each modem supports QoS queues to buffer data adequately for downlink communication. The networking and routing sub-system creates, controls, and updates routing tables to route uplink and downlink traffic between different modems.

In envisioning the future EHTS, we foresee several cutting-edge technologies and concepts being included to revolutionize SpaceNets. In order to ensure optimal allocation and utilization of satellite resources, autonomous resource management is anticipated to play a key role. A key driver in having this on the payload is latency, i.e. speed of response.  Specifically, we aim for the satellite to rapidly respond to changing user demands and provide adequate resources immediately, so the users do not notice any degradation of service quality. Doing this on the ground usually requires a few seconds, while doing it on the satellite itself can be sub-second. Therefore, it is expected that EHTS will leverage the use of Artificial Intelligence (AI)/ML techniques to provide enhanced levels of autonomy and control. In particular, AI and ML have already proven to be useful tools to accelerate complex and expensive optimization procedures for wireless networks and SatCom. Recently, the European Space Agency (ESA) initiated a project to extend AI and cloud computing into space, with the goal of demonstrating how these cutting-edge technologies could enrich future satellite missions~\cite{ESA2023}. This ambitious endeavor primarily aims to introduce greater autonomy to satellites, paving the way for the realization of smart satellite operations, all achieved without the need for human intervention.

In addition, a considerable benefit will be provided in this regard by the incorporation of a regenerative onboard multicore processor, which will guarantee dynamic and adaptive resource allocation in real-time to account for the very volatile nature of space communication. 

\begin{figure*}
  \centering
     \includegraphics[width=1\textwidth, height=5in,keepaspectratio]{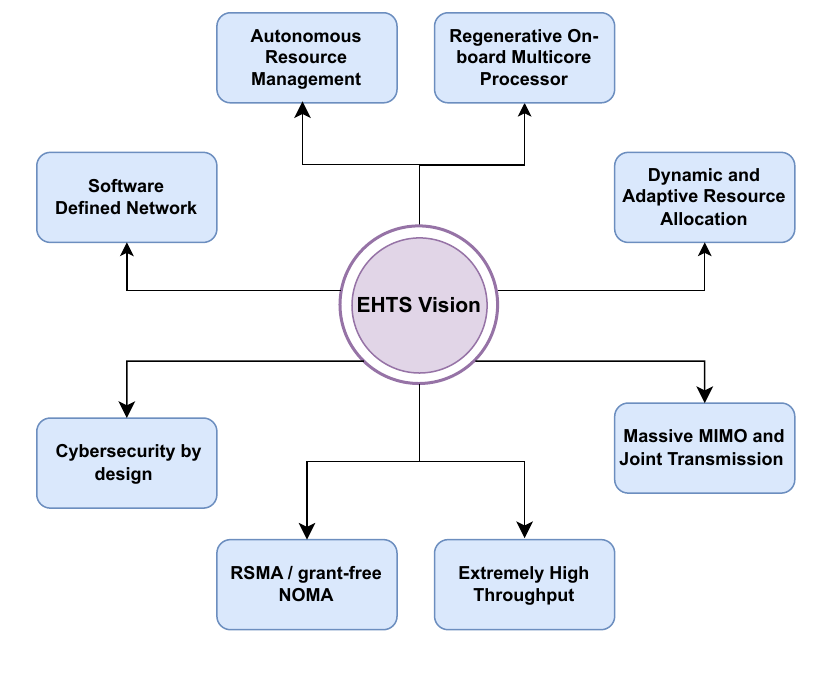}
   \caption{Main Key Technologies of EHTS}
  \label{feature}
\end{figure*}

Massive MIMO and joint transmission, two emerging ideas from terrestrial communications, are anticipated to find their way into satellite systems, increasing the system's capacity and reliability by using multiple antennas at both the transmitter and receiver ends. Along with these developments, the combination of RSMA and grant-free NOMA schemes enables efficient resource allocation without prior transmission grants, effectively serving a large number of users concurrently, even in high-demand hot-spot regions.



 Another essential component of this strategy is the adoption of cutting-edge cybersecurity measures. Strong security measures are essential given the satellite systems' growing complexity and interconnection. These would guarantee the integrity and confidentiality of sent data and aid in protecting against potential online dangers. In particular, the growing sophistication of cyberattacks on satellite communication systems, evidenced by incidents such as the hacking of Viasat during the onset of the Ukraine war~\cite{viasatattack}, has accentuated the need for robust cybersecurity in space systems. Despite the existence of some guidelines by organizations like the National Institute of Standards and Technology, critics argue that current standards lack specificity. This urgent call for secure-by-design specifications within space system components is driving a shift towards more rigorous standardization, particularly as devices and attackers alike become increasingly adept~\cite{cyberscoop2023}. Recently, the IEEE P3349 Working Group started a standard entitled \say{International technical standard for space system cybersecurity} \cite{IEEEP3349WG}. The main purpose of this standard is to establish comprehensive cybersecurity controls across the space systems ecosystem, addressing vulnerabilities in the ground systems, space vehicles, link segments, and the integration layer to ensure robust protection and trustworthiness in the ever-evolving space domain. In addition, lately, the national institute of standards and technology (NIST) has developed a cybersecurity framework for hybrid SatNets \cite{McCarthy2023}. This framework offers guidelines for organizations involved in satellite operations, ensuring designs align with their risk tolerance. It's suitable for various space-based applications with multiple stakeholders.
 

EHTS would eventually be able to attain extremely high throughput thanks to the combination of these cutting-edge technologies and concepts, making a future SpaceNets system more effective, adaptable, secure, reliable, and resilient.
The main key technologies for EHTS are summarized in Figure \ref{feature}.


\section{Conclusion }
\label{sec:Conclusion}

In conclusion, as digital technologies continue to advance and the demand for robust data connectivity rises, the role of SpaceNets becomes increasingly pivotal. The next-generation communication technologies like 6G promise transformative advancements, necessitating a more extensive exploration of non-terrestrial communication media. Recent developments in HTS and cutting-edge digital payload technologies present a promising future for SpaceNets, particularly in providing high-speed data connectivity to remote and underserved areas. Innovations such as OBP, SDN, and NFV add to this optimism by bringing adaptability and dynamism to SatNets. Despite the challenges associated with managing complex digital payloads and routing decisions, these innovations in satellite systems hold significant promise for the future SpaceNets, providing high-speed data connectivity across vast geographical regions and catering to the ever-growing digital demands of the 2030 era. 

To conclude, this paper has aimed to provide a comprehensive study of the existing literature on HTS systems, encompassing aspects of standardization, patent landscape, multiple channel access techniques, routing, scheduling, load balancing, QoS, and the importance of SDN. In addition, we present our vision for beyond HTS termed as EHTS. We also presented an overview of future regenerative payload based on multiple modem banks architecture. Through this extensive review, we have identified a significant gap in research, particularly concerning flow allocation within the satellite payload among multiple modem banks. This underexplored area represents a crucial opportunity for future research and innovation to enhance the efficiency and reliability of future HTS systems.

As research and development efforts continue, we can anticipate further breakthroughs in this domain, bringing us closer to a more connected and data-driven world.

\balance
\bibliographystyle{IEEEtran}
\bibliography{ref}
\end{document}